\documentstyle[sprocl,epsfig]{article}

\def\ra{\rightarrow}

\def\be{\begin{equation}}
\def\ee{\end{equation}}
\def\bea{\begin{eqnarray}}
\def\eea{\end{eqnarray}}
\def\ee{e^+e^-}
\def\ra{\rightarrow}

\begin{document}

\title{BEAM RELATED SYSTEMATICS IN HIGGS BOSON MASS MEASUREMENT}

\author{A.RASPEREZA}

\address{DESY, Notkestrasse 85, D-22607 Hamburg Germany}

\maketitle

\begin{center}
{\it{Proceedings of the International Conference on Linear Colliders}}\\
{\it{LCWS 2004, Paris, 19$-$23 April 2004}}
\end{center}

\vspace{5mm}

\abstracts{
The effect of beam related systematics, namely uncertainty
in the beam energy and differential luminosity spectrum 
measurements and beam energy spread on the precision
of the Higgs boson mass measurement at a future 
linear $\ee$ collider is investigated.  
}

A future linear  $\ee$ collider 
will be an ideal tool to measure Higgs boson properties. 
A number of studies have been performed  
to examine physics potential of the linear 
collider in terms of attainable statistical errors on the 
observable quantities such as Higgs boson mass, decay 
branching fractions and production rate. 
However, most of these studies did not take into
account possible systematic effects which may influence the
precision of these measurements. 
In this note possible impact of the beam related 
systematic errors on the Higgs boson mass measurement 
is discussed.  
Previous analysis has demonstrated that Higgs boson mass 
can be measured with the statistical precision of 40 MeV
for Higgs boson mass of 120 GeV~\cite{hmass}. In our 
study this analysis has been complemented 
with the simulation of the following effects: 
\begin{itemize}
\item{uncertainty in the beam energy measurement;}
\item{beam energy spread;}
\item{uncertainty in the differential luminosity spectrum measurements.}
\end{itemize}
The Higgs boson mass is measured
exploiting Higgs-strahlung process, in which Higgs boson is 
produced in association with the Z boson. 
The investigated topologies include the following 
final states: $HZ\ra b\bar{b}q\bar{q}$, $HZ\ra b\bar{b}e^+e^-$
and $HZ\ra b\bar{b}\mu^+\mu^-$. 
The analysis is performed for the centre-of-mass energy of 350 GeV
and integrated luminosity of 500 fb$^{-1}$. 
Signal and main background processes, 
$W^+W^-$, $ZZ$ and $q\bar{q}$ production, 
are generated using PYTHIA~\cite{pythia}. 
Detector response is simulated using fast parametric Monte Carlo 
program SIMDET~\cite{simdet}. 
Beamstrahlung is accounted for using CIRCE~\cite{circe}.

Analysis proceeds as follows.
Initially selection of specific final state is applied, 
exploiting event shape variables and b-tag information. 
In the case of the $HZ\ra b\bar{b}e^+e^-$
and $HZ\ra b\bar{b}\mu^+\mu^-$ channels, the presence of 
two isolated leptons, $e^+e^-$ or $\mu^+\mu^-$, with 
invariant mass compatible with the mass of Z is required.
Hadronic part of an event is then clustered into 4 or 2 jets
for $HZ\ra b\bar{b}q\bar{q}$ and  $HZ\ra b\bar{b}\ell^+\ell^-$
channels, respectively. 
Kinematical fit, imposing four momentum conservation  
is performed to improve the Higgs boson mass resolution and 
using energy and angular resolution functions for jets and leptons 
derived from Monte Carlo studies.
For the $HZ\ra b\bar{b}q\bar{q}$ channel in addition to the
four momentum conservation requirement invariant mass of the
two jets assigned for the Z boson is constrained to the Z mass.
The Higgs boson mass is reconstructed after kinematical fit as 
the invariant mass of the two jets assigned for the Higgs boson.
Higgs boson mass and corresponding statistical error are 
obtained from the fit of resulting reconstructed Higgs boson 
mass spectrum with the superposition of background and signal distributions.
Detailed description of the analysis can be found in Reference~\cite{hmass}. 

\begin{figure}[h]
\begin{minipage}[c]{0.48\textwidth}
\includegraphics*[width=0.99\textwidth]{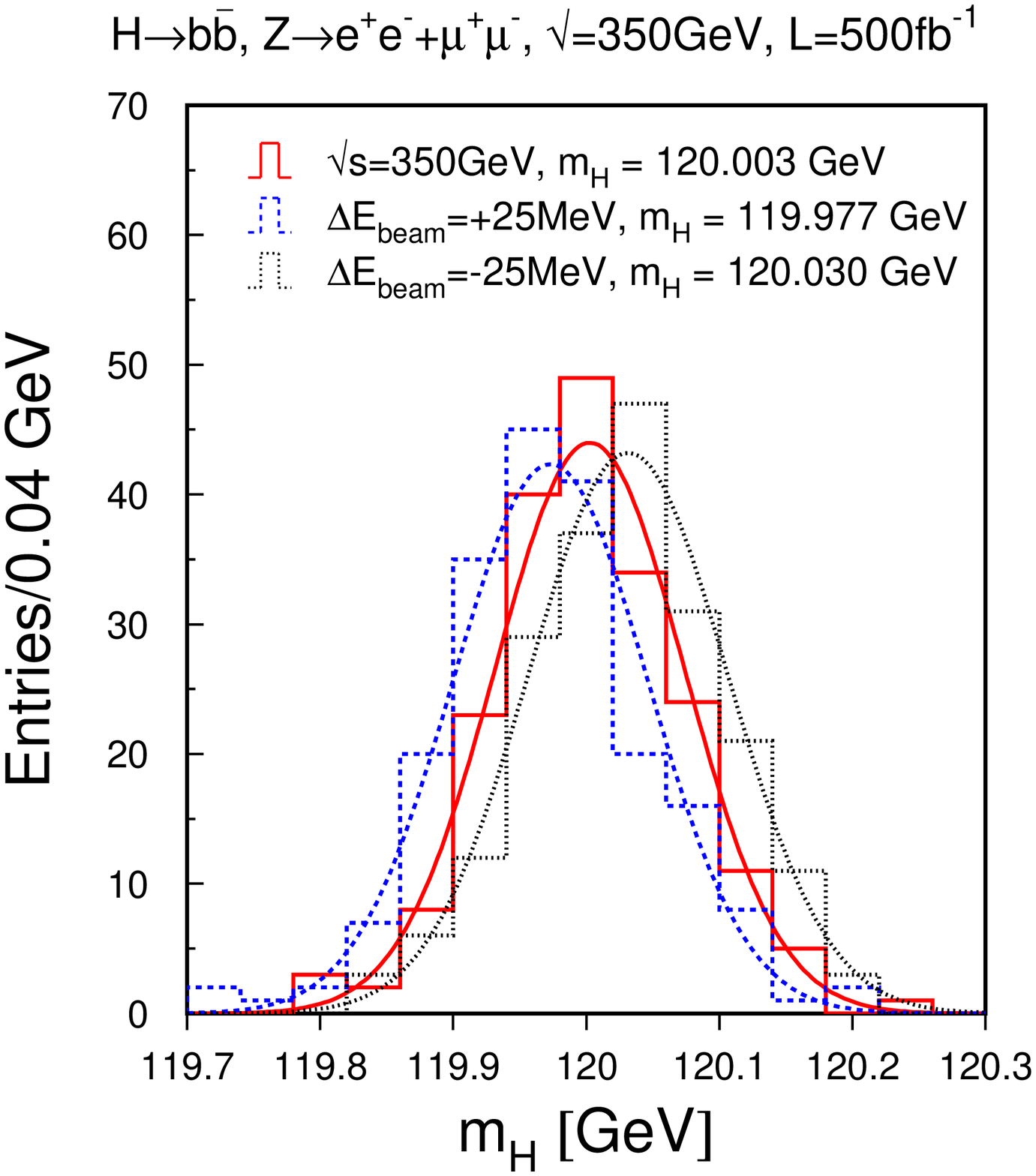}
\caption{The spectrum of the fitted values of the Higgs 
boson mass as obtained from 200 independent signal 
samples for the case when 
both electron and positron beam energies are overestimated 
by 25 MeV (dotted histogram), 
when they are underestimated by 25 MeV (dashed histogram) 
and when no shifts are introduced to the beam
energies (solid histogram).
\label{fig:beam_error}}
\end{minipage}
\begin{minipage}[c]{0.02\textwidth}
.
\end{minipage}
\begin{minipage}[c]{0.48\textwidth}
\vspace{-9mm}
\includegraphics*[width=0.99\textwidth]{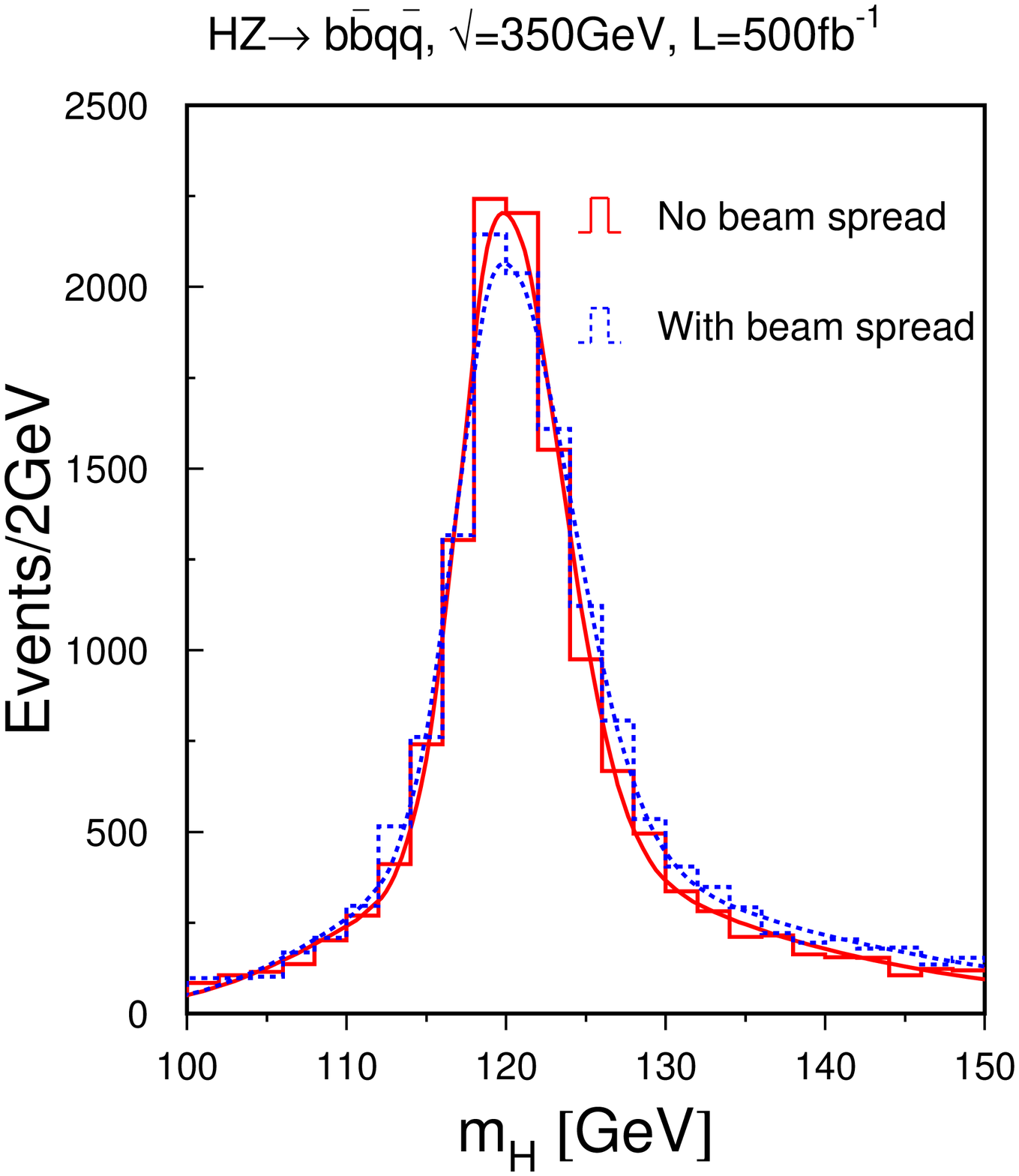}
\caption{Reconstructed Higgs boson mass spectrum in the 
sample of the $HZ\ra b\bar{b}q\bar{q}$ events for the case of monochromatic
beams (solid histogram) and for the case of 0.5\% Gaussian 
energy spread for both electron and positron beams (dashed histogram). 
\label{fig:beam_spread}}
\end{minipage}
\end{figure}

The impact of the uncertainty in the beam 
energy measurement is estimated in the following way. 
At the stage of generating signal samples
both positron and electron beam
energies are artificially shifted with respect to the nominal value
of 175 GeV. The shifts to the beam energies are varied from -100 to 100 MeV in 
25 MeV steps. Since the kinematical fit uses the nominal value
for the centre-of-mass energy of 350 GeV, the shift in the beam 
energy will result also in the shift in the measured Higgs boson mass. 
As an example Figure~\ref{fig:beam_error} 
shows the distributions of fitted values of Higgs boson mass 
in the $HZ\ra b\bar{b}\ell^+\ell^-$ channel for three scenarios : 
when both electron and positron beam energies are overestimated 
by 25 MeV, when they are underestimated by 25 MeV and when 
no shifts are introduced to the beam
energies. In each of the three considered cases the distributions 
are obtained from 200 statistically independent signal samples.
Corresponding systematic error on the Higgs boson mass is found to 
depend linearly on the uncertainty in the beam energy:
\begin{itemize}
\item{$\delta(m_H) \sim 0.8 \cdot \delta(E_b)$ for 
the $HZ\ra b\bar{b}q\bar{q}$ channel,}
\item{$\delta(m_H) \sim \delta(E_b)$ for 
the $HZ\ra b\bar{b}\ell^+\ell^-$ channels.}
\end{itemize}
At this point, one can conclude that in order to keep systematic
error at the level of statistical one, the beam energy
must be measured with relative precision of 10$^{-4}$.

To estimate the impact of beam energy spread, Gaussian smearing 
of beam energy has been applied at the stage of generating signal
events. As an example, Figure~\ref{fig:beam_spread} 
shows reconstructed Higgs boson mass spectrum in the sample of 
$HZ\ra b\bar{b}q\bar{q}$ events for the case of 
ideal monochromatic beams and for the case of 0.5\% energy spread
for both electron and positron beams. In the latter case the statistical
precision in the Higgs boson mass measurement degrades from 45 to 50 MeV
in the $HZ\ra b\bar{b}q\bar{q}$ channel and from 70 to 80 MeV in the 
$HZ\ra b\bar{b}\ell^+\ell^-$ channel.
For the TESLA machine the expected energy spread amounts to 
0.15\% for electron beam and 0.032\% for positron beam~\cite{tdr_machine}. 
In this scenario no significant degradation of the precision in 
the Higgs boson mass measurement is observed.

\begin{figure}[h]
\begin{minipage}[c]{0.48\textwidth}
\vspace{-2mm}
\includegraphics*[width=0.99\textwidth]{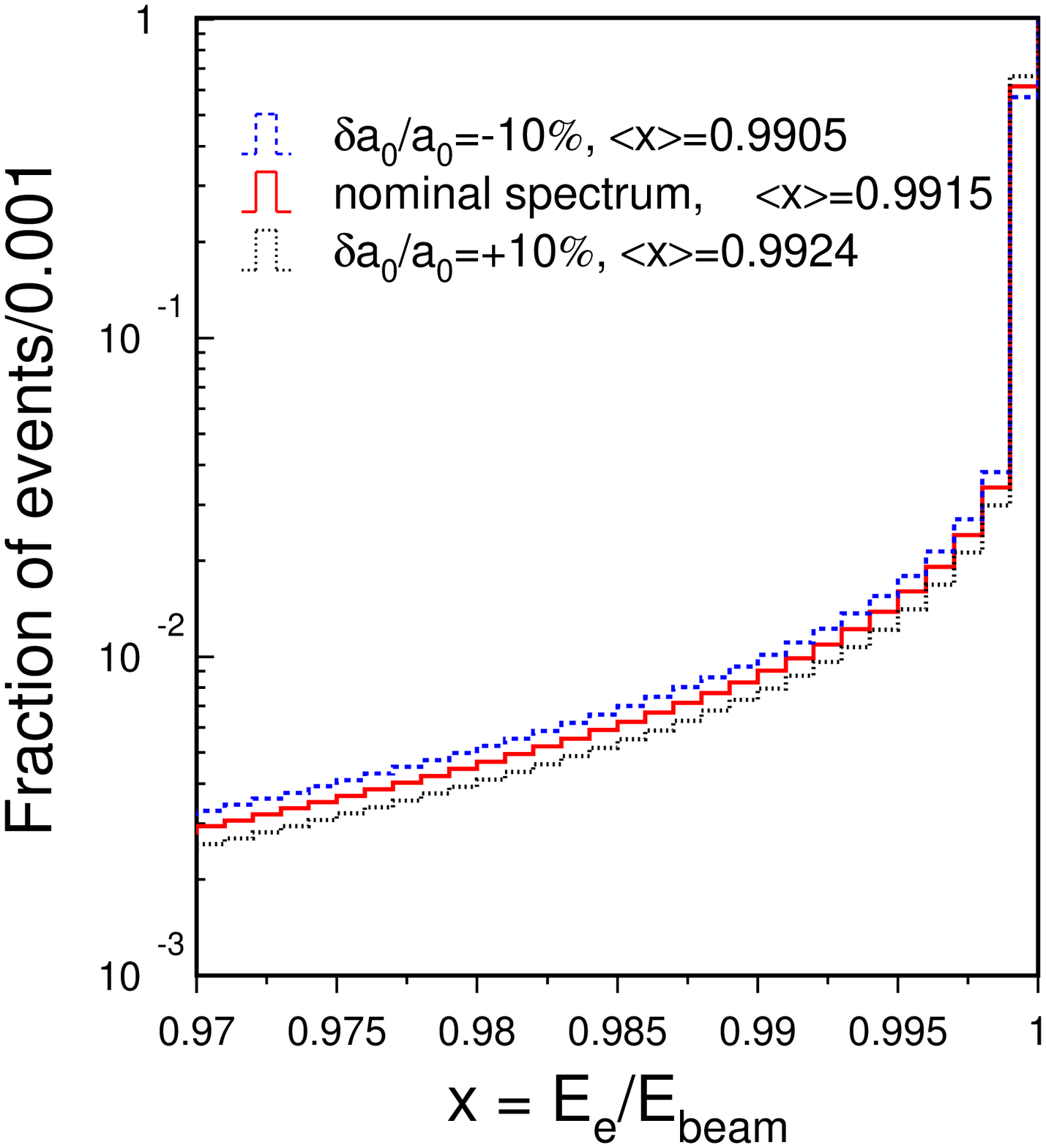}
\caption{Beam energy spectrum after beamstrahlung 
for nominal parameters $a_i$ (see text) at 350 GeV centre-of-mass energy 
(solid histogram) and for the cases when parameter $a_0$ 
is shifted from its nominal value by -10\% (dashed histogram) and +10\% 
(dotted histogram). 
\label{fig:circe}}
\end{minipage}
\begin{minipage}[c]{0.02\textwidth}
.
\end{minipage}
\begin{minipage}[c]{0.48\textwidth}
\includegraphics*[width=0.99\textwidth]{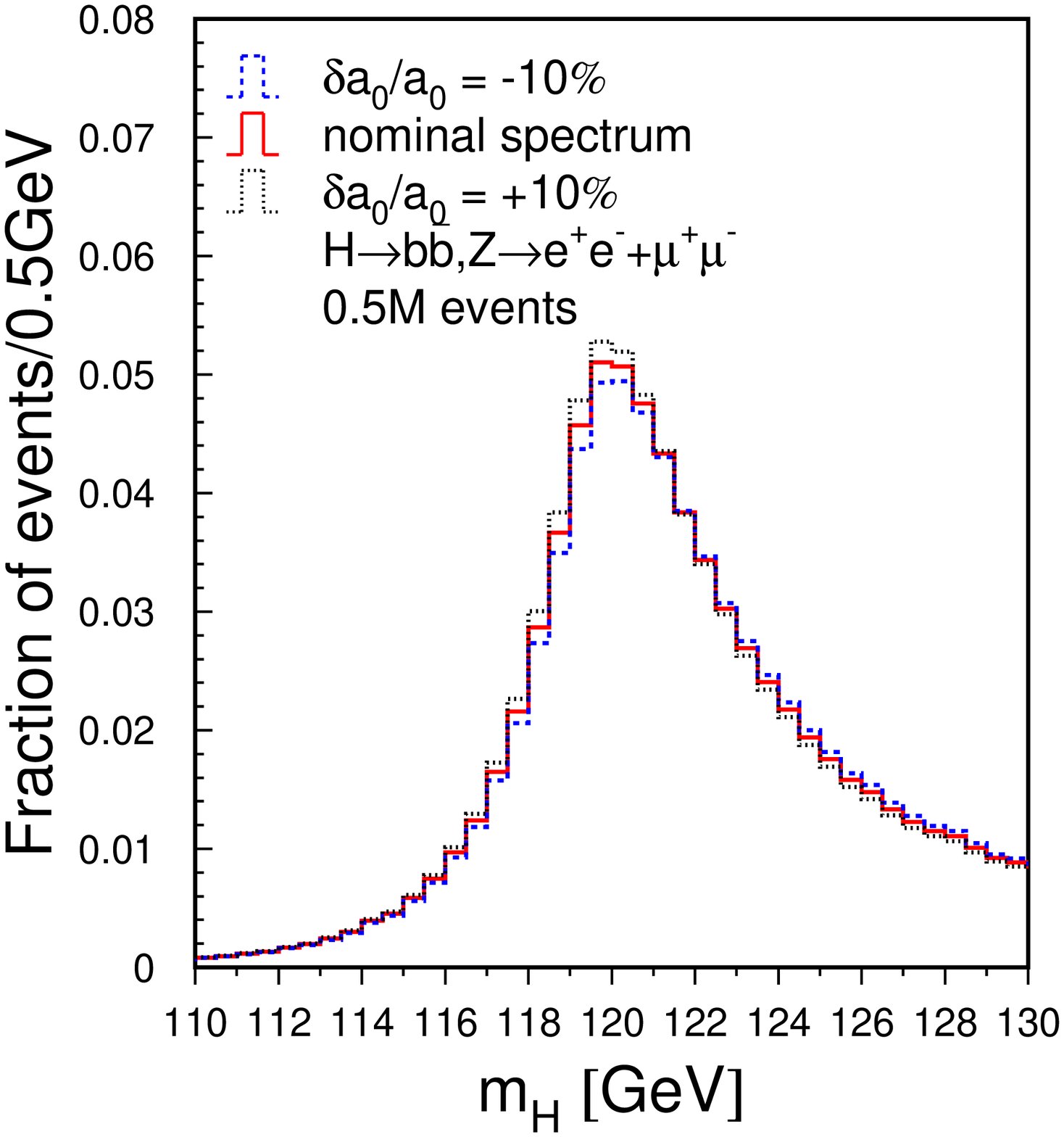}
\caption{Reconstructed Higgs boson mass spectrum 
in the sample of $HZ\ra b\bar{b}\ell^+\ell^-$ events
for nominal parameters $a_i$ (see text) at 350 GeV centre-of-mass energy 
(solid histogram) and for the cases when parameter $a_0$ 
is shifted from its nominal value 
by -10\% (dashed histogram) and +10\% (dotted histogram).
\label{fig:bstrahl}}
\end{minipage}
\end{figure}

The energy spectra of colliding electron and positron at linear 
collider will be significantly affected
by photon radiation of an electron/positron in one bunch against coherent 
field of opposite bunch. This effect is referred to as beamstrahlung.
To have a fast simulation of beamstrahlung the program CIRCE 
has been written which assumes that the beamstrahlung in 
the two beams is equal and uncorrelated between the beams and
parameterizes beam energy according to
\begin{displaymath}
f(x) = a_0\delta(1-x)+a_1x^{a_2}(1-x)^{a_3},
\end{displaymath}
where $x$ is the ratio of colliding electron/positron energy to initial
energy of undisrupted beam. Parameters $a_i$ depend on operational  
conditions of linear collider. Normalization condition, $\int{f(x)dx=1}$,
fix one these parameters, leaving only three of them independent. 
The default parameters for the TESLA machine 
operated at the centre-of-mass energy of 350GeV are:
\begin{displaymath}
a_0=0.55,\hspace{2mm}a_1=0.59,\hspace{2mm}a_2=20.3,\hspace{2mm}a_3=-0.63.
\end{displaymath}
It was shown that from the analysis of acollinearity spectrum in bhabha
events, parameters $a_i$ can be determined with a precision better 
than 1\%~\cite{lumi}.  
To visualize an effect of the uncertainty in determination of parameters 
$a_i$ in figure~\ref{fig:circe} we show distribution of beam energy spectrum
for nominal values of parameters $a_i$ and for the cases when 
parameter $a_0$ is shifted by $\pm$10\% from its nominal value.  
Figure~\ref{fig:bstrahl} presents corresponding Higgs boson mass 
spectra for the sample of $HZ\ra b\bar{b}\ell^+\ell^-$ events. 
An uncertainty of 10\% in the determination of parameters $a_i$ 
results in a systematic error of ${\cal{O}}$(10MeV) 
on the Higgs boson mass
in both $HZ\ra b\bar{b}\ell^+\ell^-$ and $HZ\ra b\bar{b}q\bar{q}$ channels.
The error is reduced to  ${\cal{O}}$(1MeV) if parameters $a_i$ are measured
with an accuracy of 1\%.

\end{document}